# Predicting Tomorrow's Headline using Today's Twitter Deliberations

Roshni Chakraborty

Comp Sc. & Engr.
IIT Patna

Bihar, India
roshni.pcs15@iitp.ac.in

Abhijeet Kharat

Comp Sc. & Engr.
IIT Patna

Bihar, India
abhijeet.mtcs17@iitp.ac.in

Apalak Khatua

Strategy Area

XLRI Jamshedpur

Jharkhand, India

apalak@xlri.ac.in

Sourav K Dandapat
Comp Sc. & Engr.
IIT Patna

Bihar, India

sourav@iitp.ac.in

Joydeep Chandra
Comp Sc. & Engr.
IIT Patna

Bihar, India

joydeep@iitp.ac.in

## ABSTRACT

Predicting the popularity of news article is a challenging task. Existing literature mostly focused on article contents and polarity to predict the popularity. However, existing research has not considered the users' preference towards a particular article. Understanding users' preference is an important aspect for predicting the popularity of news articles. Hence, we consider the social media data, from the Twitter platform, to address this research gap. In our proposed model, we have considered the users' involvement as well as the users' reaction towards an article to predict the popularity of the article. In short, we are predicting tomorrow's headline by probing today's Twitter discussion. We have considered 300 political news article from the *New York Post*, and our proposed approach has outperformed other baseline models.[1]

## CCS CONCEPTS

• Information systems---Data management systems---Database design and models---Data model extensions

## KEYWORDS

Popularity prediction of Online News; Social Media; Twitter; New York Post; Political News

## 1 Introduction

Gone are those days when an office going New Yorker used to board the subway with a folded newspaper in his hand. Reading morning newspapers on New York's subway is becoming outdated. Things have changed drastically in recent times. Today's millennial generation is not only emotionally but also physically tied to their smartphones and tablets. This has severely affected the newspaper industry. All leading newspapers across the globe have reported a sharp drop in their print circulation. Therefore, the future of this industry lies on the digital platform. The competition in this newspaper industry is not anymore about sending the print version to the remotest corner of the country. The challenge of this digital platform is to understand the latent psychological aspects of the users. If a newspaper fails to satisfy the user, then within the next few seconds she will switch to another news-related app. This will directly affect the ad revenue of a news outlet. Between various news related apps, and various social media platforms, users these days are spoilt for choice. User loyalty is a concept of a bygone era in this digital age, and the user's preferences are heterogeneous. In brief, the phenomenal growth of online news consumption and innumerable news sources has significantly increased the competition among news media outlets. In addition, the continuous influx of newsworthy events further aggravates the situation. Thus, for media outlets, the need of the hour is to develop an automated system that can help them to predict which of the today's headlines will maintain its popularity tomorrow.

Existing literature has attempted to address this. However, this stream of research mostly explored various features and contents of the articles and the title of the articles [7], [20]. Prior studies considered the subjectivity and polarity of contents [7] [9], the sentiment of the headline [30], and so on. In other words, these works focus broadly into the articulation aspects of an article. Few studies also considered the importance of an event to predict the popularity of news article [33]. One of the major shortcomings of the above approach is that hypothetically, two articles might have similar feature and polarity, but the reaction of readers might be different. It would be similar to comparing an apple to an orange even though they might be nearly similar in shape and weight. We argue that probing the social media platform can hint – which is orange, and which is apple. Social media platforms can hint about the users' preference.

[1]

Nowadays social media platforms, such as Twitter, generate an enormous amount of user-generated data, and many times this social media platforms become the mirror of the society. Existing literature has successfully explored the Twitter data to predict the election outcome [43], to understand socio-political movement [44], to study an epidemic [45] and so on. Therefore, we argue that understanding the finer nuances of Twitter deliberation can be beneficial to predict the popularity of news article on digital platforms. This paper attempts to address this research gap.

Prior studies noted that analyzing social media platform could shed light regarding the popularity [10] and the life cycle of various news article [3]. However, these studies have considered tweets, which have exclusively mentioned the URL of news article. In fact, these studies failed to probe the richness of the Twitter platform by restricting them to a very small sub-sample of tweets with the URL of a specific news article. On the contrary, our study takes a more holistic approach than these studies and considers both users' involvement with a news article and users' reaction towards a specific news article. However, the biggest challenge for this approach is to identify the relevant tweets for a specific news article. Therefore, we have developed an iterative and adaptive algorithm that considers both textual and semantic attributes to identify the relevant tweets. Our *user involvement* aspect considers various count measures, such as a total number of tweets and average number of retweets, count of hashtags, the cumulative number of unique users as well as influential users, and so on. Also, our *user reaction* indices consider linguistic aspects of the Twitter discussion, such as variances in sentiment and emotion for a particular news article. For the sake of robustness, we have considered various machine learning algorithms and an exhaustive set of baseline models based on prior studies. Our findings on the basis of 300 news items strongly suggest that patterns of Twitter deliberations can outperform other baseline models in predicting the popularity of the news articles.

## 2  Related Works

Predicting the popularity of news article is a well-researched area. However, the genesis of this research lies in the prior works on news recommender system. This research has mostly focused on the personal preference of an individual user [18]. So, understanding the user-level latent political leaning, or bias towards a certain sport or a team, can help to predict the suitable news article for an individual user. For instance, prior studies considered user's historical preferences [39], social network data [6], [1], feedbacks [32] or combination of both preferences and feedbacks [17], [15], [28], but this stream of studies struggled due to lack of adequate data. Moreover, the user's interest can vary over time, and historical data is not available for a new user. Thus, prior studies have attempted to mitigate the challenges by considering opinions of social influencers [19], topic or temporal [41] relationships between news items and users [16] or analyzing user communities [42]. These approaches yield better results in comparison to initial studies, but still, the accuracy is not satisfactory.

In comparison to the above studies, predicting the popularity of a news article is a complex task because an efficient prediction model needs to account the heterogeneity of users. More importantly, the summation of individual preference would not be the proxy for societal acceptance. For instance, it is easy to predict what a Democrat or Republican will prefer to read on the digital platform but the task becomes complex if we try to predict what political news will engage both Democrats and Republicans. So, a dominant stream of prior works focused on the content of the news article and article headline and employed machine learning algorithms [35] to predict the popularity of a news article. For instance, existing literature considered different features of an article [9], [38], such as textual [7] and temporal features, to predict its popularity. These studies noted that article content features, such as the length of the article, the time of publishing, category or genre of the article, the author of the article and so on, could predict the popularity of a news article [20].

Another set of studies also considered the linguistic attributes of an article [11, 13] or the presence of important entities within an article [31] to investigate the issue. Existing literature also noted that the headline of an article itself [12] and the polarity within the headline [30] could be important input variables for the predicting the popularity of a news article.

Existing literature also probed the event importance to predict the popularity of a news article. For instance, Setty et al. [33] ranked news articles by linking them to a chain of recent news events. Similarly, other studies tried to explore the event importance by combining articles using topic similarity from Wikipedia [24] or by considering the causal relationships [14]. However, this approach has limitations for upcoming events or for an event, which is losing relevance among readers. In these scenarios, we argue that probing users' behavioral pattern on social media platform can hint about the popularity of news article.

Some of the prior studies considered the users' behavior on a news media outlet and argued that engagement of users could predict the popularity of a news article [36], [37]. However, it is worth noting that the news media outlet represents a minuscule of digital platform readers. To the best of our knowledge, rarely any study has considered Twitter deliberation, in an exhaustive fashion, for predicting the popularity of news article. This is the research gap in the existing literature.

Popular social media platforms, such as Twitter, not only provide users an option to share their views but also allow replying or endorsing the views of others by retweeting. In other words, Twitter provides a platform for its users to engage in deliberation. Interaction of users on the Twitter platform can shed light about users' engagement with a particular news article [29]. Thus, this paper attempts to predict the popularity of news articles using Twitter data. Some of the earlier works consider initial twitter reactions [4, 27] or content and structural features [21]. However, as we have mentioned, they have only considered tweets that have news related URLs. Thus, none of the prior studies considered the richness of Twitter data. This paper not only considers the user-level involvement (by using count measures of tweets, retweets, number of unique users and other parameters) but also probes user-level reaction towards a particular news article (by understanding the various linguistic aspects of their tweets).

## 3 Data Collection

To address our research problem, we have considered the front-page political news of the *New York Post*, which is one of the most popular newspapers in the United States. It has experienced a whopping 500% growth in the last five years with 331 million page views in March 2018. Predicting the popularity of political news is the challenging in comparison to other genres of news. For instance, a news article on climate change will uniformly affect all users. Therefore, it is easy to predict the reaction of users. However, the political news might not uniformly engage and affect all users because of their ideological heterogeneity. In this study, we have considered 300 political news, from the *New York Post*, during the period July 2016 to September 2016.

We have considered the *Twitter platform* for collecting the social media data. Twitter allows free access to approximately 1% of total tweets (in a random fashion) using the streaming API. To probe our research question, we have considered the tweets related to a particular news article. Extracting the relevant tweets for particular political news is a challenging task. To address this, we have developed an adaptive algorithm that has considered both content (similar keyword mapping) features and context (same hashtag) features of tweets to extract the related tweets of a news article. Following prior studies [5], as an initial step, we have considered a set of preliminary hashtags that have threshold keywords overlap with the representative (by top 10 TF-IDF) keywords within the news article. In other words, these preliminary hashtags, which we have initially considered to crawl tweets for a particular news article, are a bag of hashtags on the basis of the article's seed tweets [5]. However, there are certain limitations to this approach because users can use multiple hashtags for a particular news article on the social media platform but not all of them might be unique to that particular news article. For instance, social media users have used multiple hashtags for the following news article titled *"GOP blasts Obama's $400M 'secret ransom' paid to Iran"* as follows: *#whitehouse, #trump2016, #chicago, #irandeal, #obamabetrayus* and so on (refer Table 1). The last two hashtags are more specific about the news article in comparison to others. Hence, we need to consider this in our data collection as well as analysis.

To address the above concern, we have collected all hashtags related to all political news articles published in the previous one month (with respect to the publication date of the article we are considering in our analysis). This process has generated a bag of hashtags. From this bag of hashtags, we have identified a set of hashtags, which were frequently used by Twitter users and labeled them as generic hashtags. Consequently, we have labeled *#whitehouse*, *#trump2016*, and *#chicago* as generic hashtags for the above article because these hashtags were used by Twitter users for other issues/news article also.

Next, we have developed an automated system for identifying hashtags specific to a news article. We have filtered out the preliminary hashtags as the hashtags those were mentioned in a tweet T, and fulfilled the threshold criteria of keywords matching with the news article [5]. We define *article-specific hashtags* as those hashtags that were mentioned in T but not in our list of generic hashtags. For instance, the article #1 (in Table 1), we have labeled *#irandeal* and *#obamabetrayus* as article-specific hashtags. Similarly, the article-specific hashtag for the news article # 4 (also in Table 1), i.e., *Obamacare hikes has families struggling to afford insurance"* was *#repealobamacare.*

To check the accuracy of this approach, we have provided around 130 news articles along with all the hashtags to three annotators. We have considered a hashtag as an *article-specific hashtag* if the majority of annotators have marked that particular hashtag as specific to that article, or otherwise labeled it as a *generic hashtag*. We observed that our proposed approach yields an accuracy of 89% in identifying an article specific hashtags. Table 1 reports a few sample news articles and corresponding article-specific *(A)* and generic *(G)* hashtags. After identifying the article specific hashtags, we use these hashtags to extract further tweets related to that news articles.

**Table 1: Political News, Related Tweets & Hashtags**

| | News | Sample Tweets | Hashtags* |
|---|---|---|---|
| 1 | GOP blasts Obama's $400M 'secret ransom' paid to Iran | **1.** It's not like Obama ever earned any money ... he gave $400 million in cash to Iran.... *#irandeal* <br><br> **2.** I strongly oppose the Raskin-supported foreign policy toward *#iran*. we must not pay ransom to a dangerous terror regime. <br><br> **3.** The *#irandeal* is nothing but a series of bribes and secret agreements that fail to prevent Iran from nuclear capability. | #whitehouse (G), #irandeal(A), #pressecretary (G), #obamabetrayus (A), #trump2016(G), #chicago(G), #cnn(G) |
| 2 | Melania Trump: I have never lived in the US illegally | **1.** What visa enabled melania trump to work in the U.S.? *#theplotthickens, #immigration, #trump* <br><br> **2.** If Melania Trump broke immigration laws, the best punishment is ... *#melaniaImmigration #nevertrump* | #trump(G), #illegalimmigration(A), #nevertrump(G), #us(G), #melaniaimmigration(A) |
| 3 | Hillary to blame for Iranian scientist's hanging, general says | **1.** Hillary Clinton reckless emails outed an Iranian nuclear scientist who was executed by Iran for treason *#neverhillary* <br><br> **2.** *#crookedhillary* server has emails discussing nuclear scientist *#executed* by iran *#shortcircuit* | #crookedhillary (G), #neverhillary(G), #shortcircuit(G), #hillary(G) |
| 4 | Obamacare hikes has families struggling to afford insurance | 1. Shhh ... You're not supposed 2 know about health insurance rate hikes until after elections! vote *#trump #repealobamacare* <br><br> **2.** *#repealobamacare* Obama will take our money | #repealobamacare(A), #trump(G), #maga(G) |

**G*– generic; A – article specific*

Next, we have extracted the user level information. In other words, we have extracted the information related to users who had participated in the political discussion related to any of these 300 news articles. We have extracted the users' name from our Twitter corpus and identified around 1 million unique users who have tweeted at least once for our sample of 300-news article. Next, we have crawled their tweeting behavior and profile-related information. In our model, we have considered the profile-related information of a user to understand whether a user is influential. If a user has more than 1000 followers and also active on Twitter platform, we have considered the user as an *influential user*. To sum up, we have considered 1.8 million tweets for our 300 news articles made by around 1 million unique users.

## 4  Proposed Model

For predicting the popularity of a news article, we have considered two categories of social media data namely, *user involvement indices* and *user reaction indices*. We argue that the popularity of a news article among the social media users (which is a proxy for digital platform readers) can be captured by analyzing the attention that a news article is receiving and the linguistic content of the discussion on the Twitter platform on the very day of its publication. In brief, the former category considers various user-level tweet statistics, and the latter employs natural language processing techniques to understand the linguistic aspects of the social media discussions. The following sections narrate how we have operationalized the involvement and reaction indices.

## 4.1 User Involvement Indices

We capture the attention of Twitter users for a particular news article by considering the user involvement through three aspects: *tweet statistics*, *user statistics* and *hashtag statistics*. Under the tweet statistics category, we have considered the *number of tweets*, the *average number of retweets* (i.e. total number of retweets divided by total number of unique tweets) and the *average number of favorites* (i.e. total number of favorites divided by total number of unique tweets) received by a particular news article on the very day of its publication. We observe that there is a significant variance in user involvement. Some news articles receive hundreds of tweets and retweets, whereas other news articles have merely received ten to twenty tweets. For our analysis, we have divided the number of tweets received by a specific news item with the highest number of tweets (in our corpus) received by a news article.

Intuitively, the number of users getting involved with a particular news article is a good predictor of the popularity of the news article. Furthermore, we note that some users get more involved with a particular news article, and they tweet multiple times in a day. However, it is worth noting that 10 tweets from 10 different users, in comparison to 10 tweets from 1 particular user, is a better proxy to gauge the popularity of a news article. On the contrary, if a user tweets, about a particular news article, for more than once, then it also indicates her high involvement with that particular news article. Hence, we have considered these finer nuances in our analysis. We have considered the fraction of users, who have tweeted more than once for a particular news article, as *affected users*.

Subsequently, it is also important to note whether a user is influential on the social media platform or not. In other words, an influential person can be an opinion leader on social media platforms. For instance, if personalities, such as Barack Obama or Donald Trump, endorse a particular news article through their personal/official twitter handle then immediately that tweet will be retweeted by hundreds of their followers. Therefore, we have considered the fraction of *influential users* as well as *authoritative users* for each news article as an input variable in our model.

**Table 2: Political News and User Involvement Indices**

| | Title of a few sample news article on $n^{th}$ day | Published on $(n+1)^{th}$ | Generic/ Hashtags: | # of Total Tweets/ | # of Total Users/ |
|---|---|---|---|---|---|
| 1 | Suicide bombing at Pakistani hospital kills at least 63 | Yes | 16/8/8 | 1319/ 42150/ 37150 | 1319/ 1240 |
| 2 | Trump to propose big tax breaks in economic plan | Yes | 20/12/8 | 208/ 44151/ 37929 | 208/ 180 |
| 3 | Trump gives Post columnist a shout-out in economic speech | No | 5/5/0 | 10/0/0 | 10/10 |
| 4 | Obama commutes sentences for record-breaking 214 prisoners | No | 13/1/12 | 13/2/0 | 13/13 |
| 5 | 'Furious' GOP leaders plot to get Trump on track | Yes | 39/8/31 | 344/ 223/ 150 | 344/ 289 |

Next, we have considered the number of article-specific hashtags on the Twitter platform as a metric to gauge the involvement of social media users. Intuitively, higher user involvement with a news article would generate a higher number of article-specific hashtags. For instance, the news article *"Trump gives Post Columnist a shout-out in economic speech"* did not generate article specific hashtags. On the contrary, the news article *"Trump to propose big tax breaks in economic plan"* has created eight article specific hashtags (refer to Table 2). Thus, we have considered the total *number of article-specific hashtags* as an input variable in our proposed model.

## 4.2 User Reaction Indices

As we mentioned earlier, natural language processing (NLP) techniques allowed us to go beyond various count-based user-level measures. We probed the linguistic content of Twitter deliberation to understand the cognitive involvement of users with a particular news article. This cognitive involvement of users can be a good proxy to predict the popularity of a news article. Therefore, we have employed sentiment and emotion analysis to gauge the user reactions towards a particular news article. We have considered two indicators to capture the user reaction namely, sentiment variance and emotion variance.

We argue that differences of opinion would lead to higher debates and discussion on the Twitter platform. For instance, most social media users would agree with a hypothetical news article such as *"Global warming would be a serious threat in the coming decades,"* and it might create a discussion but not debate. However, a hypothetical news article such as *"President Trump is failing to take appropriate policy measures to control global warming"* would probably lead to a debate between the Democrats and Republicans. Republican will try to discard this view, and vice versa. Consequently, the discussion regarding this particular news article will gain momentum. We are attempting to capture this in our proposed model.

Following Vader Sentiment Analyzer [8] and TextBlob [22], we have calculated the average sentiment score of a tweet. We have identified all tweets specific to a particular news article and classified whether the tweet is positive or negative. Next, we have considered the *sentiment variance* of all tweets related to a particular news article to understand the differences in opinions. We have calculated the sentiment variation *(SV)* as follows:

$$SV = 1 - \{(|PC - NC|) / (PC + NC)\}$$

*PC* is the number of positive tweets for a news article, and *NC* is the number of negative tweets for the same news article. The sentiment variance will be highest when the count of positive tweets and negative tweets are equal for a news article, and the sentiment variance decreases when there will be higher number of positive/negative tweets (refer Table 3 for a few representative examples). An equal number of positive and negative sentiment indicates that users are from two ideologically opposite camps. On the contrary, only positive or negative tweets indicate that users are ideologically homogenous.

Next, we have also considered the emotional content of a tweet. We employ the NRC emotion lexicon [25], [26] to classify a tweet among various emotion classes such as anger, anticipation, trust, disgust, fear, joy and surprise. Similar to sentiment variance, we have classified all tweets specific to a particular news article into various categories of emotions. Intuitively, high emotional variance indicates that users are displaying different emotions towards a news article. We have calculated the emotional variance *(EV)* as follows:

$$EV = \{\sum_i (e(i) - m(e))^2\}/n$$

In the above formula n is the number of emotion categories which is 8 [25], [26]. *e(i)* is the fraction of tweets with $i^{th}$ emotion, and the value of *m(e)* is *(N/8)* where *N* is the total number of tweets for a particular article, and the final value (of EV) is the summation of the above measure across all emotion classes. Here, a high emotion variance indicates that tweet corpus for a particular news article represents multiple emotion categories. For instance, in response to the immigration issue related news, a Republican, who believes that strong immigration law would protect American jobs, might display joy. On the contrary, a social activist, who thinks otherwise, might display her anger to the same news article. Refer Table 3 for a few representative examples of the same.

**Table 3: Political News and User Reaction Indices**

|   | Title of a few sample news article on the $n^{th}$ day | Published on the $(n+1)^{th}$ day | Sentiment Variance | Emotion Variance |
|---|---|---|---|---|
| 1 | Suicide bombing at Pakistani hospital kills at least 63 | Yes | 0.67 | 43.19 |
| 2 | Trump to propose big tax breaks in economic plan | Yes | 0.90 | 20.30 |
| 3 | Trump gives Post columnist a shout-out in economic speech | No | 0.00 | 0.00 |
| 4 | Obama commutes sentences for record-breaking 214 prisoners | No | 0.00 | 0.00 |
| 5 | 'Furious' GOP leaders plot to get Trump on track | Yes | 0.37 | 41.51 |

## 5 Data Analysis

### 5.1 Preparation of Gold Standard

We are trying to predict whether a particular news article of the $n^{th}$ day will be followed by another subsequent article on $(n+1)^{th}$ day. However, on $(n+1)^{th}$ day the title of the subsequent article can differ significantly from the previous day. For instance, on $n^{th}$ day, the hypothetical title of a news article can be *"Why Brexit matters for the American Corporate Sector?"* However, on the $(n+1)^{th}$ day the issue will continue, but the title can be: *"American Corporates are reluctant to invest in the UK."* Therefore, it requires a contextual understanding to prepare the database for our analysis. We have employed three annotators and provided them with a particular news article of $n^{th}$ day and all the news articles of $(n+1)^{th}$ day for manual annotation. We have asked our annotators to mark a news article either as 1 if the same news is covered on the subsequent day, and 0 otherwise. For our analysis purpose, we have considered the labeling by the majority of the annotators.

### 5.2 Baseline Models

As we discussed, a plethora of studies have tried to predict the popularity of news article [11], [12], [13], [30], [31]. However, this stream of literature can be broadly classified into five categories as follows: article content and polarity, title content and polarity, and event importance categories. See Table 4 for a detailed feature list or input variables for these approaches. We have considered all these five approaches as our baseline models.

Following the prior studies [7], [20], [9], we have extracted both the content and polarity of the article to predict whether the article will be published on the next day or not. Prior studies have considered the overall sentiment of the article, usages of positive or negative words within the article, the length of the article, and so on. Similarly, we have also considered the content and polarity of the title of the article to predict its popularity. We have noted certain differences in terms of the number of features in prior studies. However, in our analysis, we have tried to consider an exhaustive set of features for the first four-baseline models namely, article content and polarity, title content and polarity. The final baseline model is the event importance [33]. This tries to capture the dominance of a topic/issue in comparison to others. So, the event importance of a news article is calculated by the numbers of similar articles, which get published on consecutive days. A pair of news articles is considered as similar to each other - if it crosses the threshold value in terms of the list of entities and bi-grams features between two articles. Prior studies noted that event importance is also an indicator to predict the popularity of a news article.

**Table 4: Baseline models vis-à-vis proposed model**

| Baseline Models | List of Features/Input Variables |
|---|---|
| Article Polarity [20] | Polarity score of the article; the rate of positive & negative words per 100 words; the rate of positive & negative words per 100 words with non-neutral words, the average polarity of positive & negative words; min. & max. the polarity of positive & negative words |
| Article Content +Article Polarity [7, 9, 20] | # of words in the article; the rate of non-stop words; day of the week on which it got published; published on weekend or not; # of entities in the news article; average word length of the article |
| Title Polarity [30] | Polarity score of the title; the rate of positive & negative words in the title; the rate of positive & negative words with non-neutral words in the title; average polarity of positive & negative words in the title; min. & max. the polarity of positive & negative words in the title |
| Title Content + Title Polarity [30] | # of words in the title; the rate of non-stop words in the title; # of entities in the title; the average word length of the title |
| Event importance [33] | # of days a news article related to the event was published, # of articles of the event was published |
| **Proposed Model** | *User Involvement Indices:* # of tweets, average # of retweets, average # of favorites, the percentage of affected users, the percentage of influential users, # of article-specific hashtags<br><br>*User Reaction Indices:* Sentiment variance, Emotion Variance |

## 6 Results

We have employed Random Forest Classified (RFC), Support Vector Machine (SVM), and Classification and Regression Trees (CART) algorithms for our analysis. We have applied these four three classifiers on our article dataset for the five baseline models as well as our proposed models. We have considered 10-fold cross-validation for our analysis. We have repeated our experiments multiple times and found our results are consistent. We have reported the same in

Table 5. Our proposed model has outperformed all five-baseline models for all three classifiers. Our F score for SVM and RFC classifiers are marginally better than the CART classifier. Broadly, the SVM classifier has outperformed other classifiers not only for our proposed model but also for baseline models.

**Table 5: Comparison of baseline vis-à-vis proposed model**

|  | RFC | SVM | CART |
|---|---|---|---|
| *Proposed Model* | | | |
| Precision | 91.4 | 94.6 | 83.3 |
| Recall | 84.2 | 83.33 | 85.7 |
| F-score | **87.6** | **88.6** | **84.9** |
| *F-score of Baseline Models* | | | |
| Article Polarity | 86.7 | 84.1 | 81.7 |
| Article Content + Polarity | 86.8 | 86.9 | 74.9 |
| Title Polarity | 85.3 | 84.9 | 83.5 |
| Title Content + Title Polarity | 86.0 | 87.8 | 82.5 |
| Event Importance | 81.5 | 87.4 | 84.0 |

## 7   Conclusions

The advent of information and communication technology has affected the newspaper industry severely in last few decades. Seamlessly connected various communication channels are generating a huge volume of information. Moreover, the digital platform is becoming a crowded place. Multiple news outlets are struggling to grab a larger share of this platform. Therefore, selecting a potentially popular news article is becoming a daunting task for the media outlets. Hence, we need to develop an automated system, which can efficiently select the news article that will most likely draw the maximum attention of users on the digital platform. To address this, prior studies focused mostly on the content and polarity of the news article to predict the popularity of news articles. However, these studies failed to capture the latent psychological aspects of users. Thus, our proposed approach is trying to gauge the users' perception from the social media discussions. We have considered the Twitter platform for our study. Our proposed model has incorporated users' involvement and reaction towards a particular news article. In short, as our title suggests that we are trying to predict tomorrow's popular headline by considering today's discussion on the Twitter platform. We have employed various machine-learning algorithms to test the accuracy of our proposed approach. We observe that our proposed approach ensures higher accuracy in comparison to other baseline models. Considering Twitter discussion for predicting the popularity of news article is the core contribution of this study.

However, there are certain shortcomings of our proposed model which future research needs to address. Firstly, we have considered a small sample of 300-news article for a relatively shorter period. Future studies in this area should consider a larger sample and longer time horizon. Secondly, we have considered the political news of the *New York Post*. In other words, we have tested our proposed approach in the political sphere of the United States. So, future studies need to probe the efficacy of our model for other genres of news in other countries. The biggest challenge will be to extrapolate this approach to a context where native language is not English. Finally, we have considered a few fundamental machine-learning algorithms. Future studies need to consider advanced deep learning based models to test the accuracy of our approach.